\documentstyle[aps,12pt]{revtex}

\begin{document}
\title{Ballistic Magnetoresitance over 4000\% in Ni-Ni Electrodeposited Nanocontacts}
\author{Hai Wang, H. Cheng and N. Garcia*}
\address{Laboratorio de Fisica de Sistemas Pequenos y Nanotecnologia}
\maketitle

\begin{abstract}
This paper reports ballistic magnetoresistance values over 4000\% measured
in electrodeposited Ni-Ni nanocontacts with T geometry previously developed
. Over the time, after several magnetic field cycles, the ballistic
magnetoresistance relaxed to a 400\%. While that the magnetoreistance of a
contact could rise indefinitely; relaxtion and reproducibility are, however,
the main issue. We find that the tip ending radius conforming the contacts
appears not to play the main role.


* corresponding author

email: nicolas.garcia@fsp.csic.es

Tel: \ 34-91-561 88 06

Fax: 34-91-563 15 60
\end{abstract}

\bigskip \newpage\ \ \ \ \ Ballistic magnetoresitance (BMR) of 300\% was
first reported by Garcia, Mu\~{n}oz and Zhao, in atomic contacts of
resistances larger than 1000 $%
\mathop{\rm %
\Omega}%
$ \cite{1} and was explained as resulting from scattering with very sharp
magnetic domain walls \cite{2,3}. Later on it was possible to obtain, by
electrodeposition techniques, contacts of a few Ohms resistance with a $T$
shape configuration (see Fig.1) having BMR of 700\% \cite{4}, that relax to
a final 
\'{}%
equilibrium%
\'{}%
value of 400\% after field cycling the sample during a few days. Contrary to
the case of atomic size contacts, the magnetoresistance of these few Ohms
contacts (10-30 $%
\mathop{\rm nm}%
$ size) cannot be attributed \cite{5} to domain wall scattering \cite{2,3}.
This has been reported by us in Ref. 4. We postulated that the physical
cause of BMR in the large nanocontacts was the presence of a dead magnetic
layer at the contact point. Thus if the polarization at the Fermi level
could be modified to a value near unity, then the resulting BMR could rise
up to infinity.

Recently BMR values over 3000\% in a Ni-Ni nanocontact with $T$ geometry
have been reported \cite{6}. The large increase of BMR over the values
previously obtained by Garcia $etal.$ \cite{4} in similar contacts (3150\%
against 700\%) was attributed to the fact that the tip ending had been
sharpened to 40 $%
\mathop{\rm nm}%
$ by electrochemical etching in KCl \cite{6} (see Fig.1).

In this communication, we report Ni-Ni contacts made by electrodeposition
with the same $T$ shape configuration that have 4000\% BMR. Both the tip I
and the wire II (Fig. 1) are treated electrochemically with KCl, but the tip
end was 1000 $%
\mathop{\rm nm}%
$ in size, and not 40 $%
\mathop{\rm nm}%
$ size as in the experiment above \cite{6}, which implies that the increase
of BMR is not due to the tip size but depends, instead, on the
electrochemical processing.

We arranged the Ni wires in a $T$ configuration as shown in Fig.1. The
applied field during magnetoresistance measurements is in the direction of
the Ni wire labeled 
\'{}%
I%
\'{}%
in Fig.1. This arrangement, originally conceived by us, is well suited for
magnetoresistance across the Ni nanocontact. The tip of the wire 
\'{}%
I%
\'{}%
in Fig.1 was positioned to within a few microns to few tens of microns to
the Ni wire labeled 
\'{}%
II%
\'{}%
prior to electro-deposition of the Ni nanocontact. The Ni wires (except for
the region in the immediate vicinity of the contact) were insulated by a
fast curing resin epoxy in order to limit the deposition to the gap between
the wires. The resin epoxy also firmly holds the Ni wires to the underlying
dielectric substrate. The Ni nanocontacts were electrochemically deposited
at room temperature. Electro-deposition was performed from a saturated
nickel sulphate (Ni$_{2}$SO$_{4}$) electrolyte (PH=3.5). We used a cathode
potential of -1.2 $%
\mathop{\rm V}%
$ against a saturated calomel electrode. Deposition times were typically
less than one minute. The magnetoresistance measurements were performed at
room temperature for magnetic fields up to H=3000 Oe in the current-in-plane
/ field-in-plane(CIP/FIP) confiuration. The process of fabrication of the
tip 
\'{}%
I%
\'{}%
was as follows: we first made a tip by breaking mechanically a Ni wire of
diameter (0.125 $%
\mathop{\rm mm}%
$). Then we used an electrochemical etching technique to get the tip radius
down to 600-1000 $%
\mathop{\rm nm}%
$. The Ni wire was deeped into a cell filled with 2M KCl and a 2$%
\mathop{\rm V}%
$ constant voltage was applied. The tip is shown in Fig.1a. Etching takes
place according to the anodic reaction $Ni(s)+2Cl^{-}=NiCl_{2}+2e^{-}$; the
reaction occurring at the Platium cathode is $2H_{2}O+2e^{-}=H_{2}+2OH^{-}$.
It should be stressed that electrochemical sharpening of the tip was not
performed, contrarily to the case in Ref. 6. It would appear that sharpening
the tip to 40 $%
\mathop{\rm nm}%
$ is not the determining step for obtaining contacts with high BMR values,
as stated in Ref. 6. We show here that one can get 4000\% BMR with a tip
sharpened to 1000 nm diameter. Instead, we believe that the additional
electrochemical treatment in the KCl cell is the cause for the high BMR.

Fig.2 shows consecutive magnetoresistance loops in a sample whose initial
zero-field contact resistance was 15 $%
\mathop{\rm %
\Omega}%
$\ after electro-deposition. This contact resistance R determines the
diameter d=$\sqrt{1000/R(%
\mathop{\rm %
\Omega}%
)}$ (in $%
\mathop{\rm nm}%
$) \cite{4} of the nanocontact, being equal to 8 $%
\mathop{\rm nm}%
$ for this sample. As seen in Fig. 2a, the resistance increases rapidly with
increasing the field strength. At saturation the resistance rises to 634 $%
\mathop{\rm %
\Omega}%
$, after which it remains essentially unchanged with further increases in
field strength. This represents a 4000\% BMR at room temperature in a field
about 800 Oe. The peak of the BMR curve corresponds to about 280 Oe. Fig. 2b
shows the BMR curve of the same sample measured after several hours. It can
be seen that, although the low field resistance increases somewhat and the
BMR value decreases significantly, the high field saturation resistance
remains virtually unchanged. This relaxation process needs to be understood
and is very important, cannot be neglected, as was already stressed in our
previous work \cite{4,7} .

In conclusion we have reported BMR over 4000\% and have shown that the size
of the tip conforming the contact does not play a relevant role because tips
1 micrometer size, as reported here, give also BMR values much larger than
the ones previously reported in \cite{6} . In fact, there is not a single
insight in the physics of the BMR process that would point to the importance
of sharpening the tip to achieve very high BMR values. The whole thing is
regulated by the nanocontacts and on the products O, S, Cl, etc, or density
of states at Fermi, that can be segregated at the contact. This has been
reported in a very recent paper (cond-mat/0207323) \cite{8} indicating that
the localization, through bonds, of the Ni sp-electrons can lead to full
polarization of d-electrons contributing to current and therefore to an
indefinitely grow of the BMR.Notice that \cite{2,3}:

\ \ \ \ \ \ \ \ \ \ \ \ \ \ \ \ \ \ \ \ \ BMR(\%)= D$\times $2P$^{2}$/(1-P$%
^{2}$)$\times $100 \ \ \ \ \ \ \ \ \ \ \ \ \ \ \ \ \ \ \ \ \ \ \ \ \ \ \ (1)

Where D is the dynamical factor approximately equal to unity for very thin
domain wall \cite{2,3} fixed by the dead magnetic layer. This is the factor
that gives the spin conservation in the current or nonadiabaticity. P is the
polarization of the electrons contributing to the current if the sp
electrons in the case at hand of Ni are bonded by Cl this will not
contribute to the current and only the fully polarized d-electrons will
participate to the current and thus P\symbol{126}1. Inserting these values
in equ. 1 it can be seen that BMR can grow indefinitely (see Ref. 9).

We thank Prof. J. A Rausell for reviewing the manuscript and M. Mu\~{n}oz
for discussions. This work has been supported by the Spanish DGICyT.

\bigskip \newpage

$\bigskip $

$References:$

\newpage

$\bigskip $

$FigureCaptions:$

Fig.1 (a) Optical microscope view of the tip conforming the T contact in
1(b). The tip and the wire have been given an electrochemical treatment in
KCl (see text). This seems to enhance BMR to very large values, irrespective
of tip ending dimensions.

Fig.2 (a) BMR cycles showing values up to 4014\%. However, a new cycle was
done after a time and the BMR was reduced to 354\% (see 2(b)). This happens
always to our samples and is the important point to understand \cite{4,7}.

\end{document}